\documentclass[runningheads]{llncs}
\def \VersionWithComments{}
\usepackage[T1]{fontenc}
\usepackage[right]{lineno}

\usepackage{graphicx}
\usepackage{packages}
\usepackage{macros}
\usepackage{alltt}     
\usepackage{hyperref}

\usepackage{color}

\begin{document}

\title{Timed  Strategies for Real-Time Rewrite Theories}

\author{Carlos Olarte\inst{1}
  \and
  Peter Csaba {\"O}lveczky\inst{2}
}

\institute{LIPN, CNRS UMR 7030,
    Universit\'e Sorbonne Paris Nord, France \\
 \and Department of Informatics, University of Oslo, Oslo, Norway}


\maketitle

\begin{abstract}
  In this paper we propose a language for conveniently defining a wide
  range of execution strategies for real-time rewrite theories, and
  provide Maude-strategy-implemented versions of most Real-Time Maude
  analysis methods, albeit with user-defined discrete and timed
   strategies.  We also identify a new time sampling strategy that
   should provide both efficient and exhaustive analysis for many
   distributed real-time systems. We exemplify the use of our language
   and its analyses on a simple round trip time protocol, and compare
   the performance of standard Maude search with our
   strategy-implemented reachability analyses on the CASH scheduling
   algorithm benchmark.
   \end{abstract}

\section{Introduction}

Real-time systems can naturally be defined in rewriting
logic~\cite{Mes92} as \emph{real-time rewrite
  theories}~\cite{OlvMesTCS}. In such theories,  actions that can be assumed to take
zero time are modeled
by ordinary (also called \emph{instantaneous}) rewrite rules, and time
advance is modeled by labeled ``tick'' rewrite 
rules of the form  $[l]\!:  \texttt{\char123}t_1\texttt{\char125} 
\longrightarrow \texttt{\char123}t_2\texttt{\char125} \mbox{ \bf in
  time } \tau  \mbox{ \bf if } \mathit{cond}$, where the \emph{whole}
system state has the form \texttt{\char123}$t$\texttt{\char125}.

Real-time rewrite theories inherit the expressiveness and
modeling convenience of rewriting logic, and allow us to  model
a wide range of  distributed real-time systems---with
different communication forms, user-defined data types, dynamic object
 creation and 
deletion, and so on---in
an object-oriented style.

The specification and analysis of real-time rewrite theories is
supported by the Real-Time Maude~\cite{rtm-journ,tacas08,wrla14}
language and tool, which is implemented in Maude as an extension of
Full Maude~\cite{maude-manual}.
For dense time, the tick rules typically have the form
\texttt{crl [tick] : \char123\(t\){\char125} => \char123\(u\){\char125} in
  time T if T <= \(f(t)\)}, where \texttt{T} is a variable of sort
\texttt{Time} not appearing in the term $t$~\cite{rtm-journ}.

Real-Time Maude provides \emph{explicit-state} analysis
methods, where the above tick rules are executed according to a 
\emph{time sampling strategy}, where the variable \texttt{T} in the
rule above is instantiated either to:
\begin{itemize}
\item a user-selected value (such as 1); or to
  \item the maximal possible time increase $f(t)$ (or a 
    value $\Delta $ if $f(t)$ is infinity).
  \end{itemize}
Real-Time Maude supports unbounded and time-bounded reachability
analysis, LTL and timed CTL model checking, and other
time-specific analyses. 
All such
explicit-state analyses are performed with the selected time sampling
strategy.

The cost of the expressiveness and generality of Real-Time Maude is
that  time-sampling-based analysis is not sound in general
for dense time~\cite{wrla06}.
Nevertheless, counterexamples/reachable
states obtained by Real-Time Maude analysis are real
counterexamples/reachable states. Real-Time Maude has therefore been
used to discover subtle but significant bugs in a number of sophisticated
state-of-the-art systems beyond the scope of decidable formalisms like
timed automata, including: a 50-page active network
protocol~\cite{aer-journ} (which required advanced functions and
detailed modeling of communication), state-of-the-art wireless sensor
network algorithms~\cite{wsn-tcs} (which required modeling coverage
areas, angles, etc., and functions on such), mobile ad-hoc network
leader election protocols~\cite{manet-journ} (the fault was due to a
subtle interplay between node movements and communication delays),
scheduling algorithms with reuse of unused budgets~\cite{fase06}
(which somewhat surprisingly required unbounded queues), a traffic
intersection system from the Ptolemy II
library~\cite{ptolemy-journ,lepri-journ} (which required defining the
semantics of Ptolemy II discrete-event models), cloud-based
transaction systems~\cite{kokichi,cloud-chapter}, and an error in running cars
 which was not found by other methods, and where
reportedly Real-Time Maude time sampling was  key.\footnote{Hitoshi
  Ohsaki, personal communication, 2007.}

Real-Time Maude's expressiveness  and generality also makes it a
suitable  semantic
framework and formal analysis backend in which (subsets of) modeling languages such
as  AADL~\cite{fmoods10}, Ptolemy II DE
models~\cite{ptolemy-journ}, Timed Rebeca~\cite{timed-rebeca-journ},
and a DoCoMo 
Labs handset language~\cite{musab-l} have been given a formal semantics
and formal analysis 
capabilities~\cite{carolyn}.

 Maude was recently 
equipped with a strategy language to allow users to define
specific  execution strategies on top of their Maude
specifications~\cite{maude-strategy-language}. In this paper we use Maude's strategy
language to define useful strategies for real-time rewrite
theories. This work is motivated by the following issues:
\begin{enumerate}
    \item Real-Time Maude analyses  apply 
       one of the above   time sampling
      strategies to \emph{all} applications of tick rules. However, as the
      following  example 
      shows,  more sophisticated  time sampling strategies
      are often desired: 

      Consider a  system 
          computing the \emph{round trip time} (RTT) between two
        nodes every five seconds. 
        A time sampling strategy that
        visits every time unit  covers all possible behaviors in
        discrete time domains, but  visits  each time point
        even when the round trip time in that round  already has been
        found, leading to \emph{very inefficient}  analyses.  On
        the other hand, increasing time maximally in each tick step  only takes into
        account those behaviors where each message has been delayed as much
        as possible, thereby  always finding the same (maximal) round
        trip time value,
        which is  not very helpful.

        In this simple but prototypical example, the ideal time
        sampling strategy  advances  time by one time unit as long
        as there is a message (which could arrive at ``any'' time) in
        the state, and  increases time maximally when there is no
        message in the state (and we are just idling until the next
        iteration of the RTT protocol). Such time sampling would 
         cover all possible behaviors, yet
        would not stop time unnecessarily.   


      \item 
        The user may also want to define her own
        \emph{non-time-sampling} (called \emph{discrete}) 
        execution strategies  (e.g., what happens if we
         restrict the model of a coffee machine so that the user can
         add at most two spoons of sugar?). We 
        should therefore support the analysis of real-time systems
        with user-defined execution strategies on the non-time parts,
        combined with the desired (possibly 
        user-defined) time sampling strategy. This includes typical 
        ``timed strategies'' in
          real-time systems including: 
           \emph{eagerness} of all/some actions; given priority to some actions;
          deciding the next action according  to the previously states
          visited;   and so on. Such more advanced strategies cannot be easily 
          defined on top of the current implementation of Real-Time Maude.

        \item Maude's strategy language  has an efficient implementation,
          using multi-threading  and the option of
          depth-first search analyses, which  provides better
          performance than standard Maude search in some cases, as shown in  Section~\ref{sec:examples}.

\item Real-Time Maude is  implemented as an extension of  Full 
    Maude. While this allowed us to use Full Maude's infrastructure to
    define distributed real-time systems in a very useful
    object-oriented style, extending the large Full Maude
    implementation posed challenges in terms of  integrating other
    analysis methods for real-time rewrite theories into Real-Time
    Maude, and of maintaining and upgrading the tool to  newer
    versions of  Full Maude.   Since Maude 3.3 directly supports
    object-oriented specification, and since Full
    Maude  will no longer be maintained and
    upgraded, we are currently working on developing the next version
    of Real-Time Maude as a much ``lighter'' Maude implementation that
    does not extend Full Maude. In this context, doing as much as
    possible as easily as possible using available Maude features is
    needed.    
              \end{enumerate}

In this paper we therefore show how most Real-Time Maude analysis
methods can be performed by rewriting with strategies directly in Maude
(Section~\ref{sec:direct-maude-strat-lang}).

However, even those  analysis methods needed somewhat
hard-to-understand strategy expressions, even without  advanced time
sampling strategies.   This begs the question how
the casual Maude user can 
analyze her system with more complex discrete and time sampling
strategies. Furthermore, since (Real-Time) Maude can be used as a
formal analysis backend for many real-time modeling languages, we
really need  an intuitive way to define useful strategies for
the non-expert Maude user.   For example,
in~\cite{ftscs-journal,DBLP:conf/apn/AriasBOOPR23} we 
claim as one of the main selling point of our Maude framework, compared to
state-of-the-art domain-specific tools for parametric timed automata
and time Petri nets,  the ability to analyze the system with user-defined
execution strategies. However, this selling point becomes moot if the
timed automata/Petri net expert cannot define her strategies.

To address this issue, in Section~\ref{sec:core-strat} we define and
implement what
we hope is an intuitive and  fairly powerful timed strategy language for real-time
rewrite theories. This language  should make it easy for the casual user to define a
wide range of useful discrete strategies as well as advanced  state- and
even history-dependent time sampling strategies in a modular way.

Even the most intuitive language would  be useless if systems
cannot be analyzed efficiently.   In Section~\ref{sec:examples} we
therefore compare the performance of
standard Maude search with our strategy-implemented analysis methods on a
sophisticated scheduling system~\cite{fase06}.   
It turns out that using a depth-search strategy on this system
sometimes  leads 
to a faster analysis.

Finally, we discuss related work  in
Section~\ref{sec:related}, and give some concluding remarks in
Section~\ref{sec:concl}. The strategy language and its implementation,
with Maude models and their strategies and execution commands
are available at \cite{rt-strategies}.

\section{Preliminaries}  \label{sec:prelim}

\paragraph{Rewriting Logic and Maude.}

Maude~\cite{maude-book} is a  rewriting-logic-based executable formal
specification  language and
high-performance analysis tool for  distributed systems.
%
A Maude module specifies a 
 \emph{rewrite theory} $(\Sigma, E\cup A,  R)$, where:
\begin{itemize}
\item $\Sigma$ is an algebraic \emph{signature}; i.e., a set of 
\emph{sorts}, \emph{subsorts}, and \emph{function
    symbols}.  
\item $(\Sigma, E\cup A)$ is a \emph{membership equational
logic}~\cite{membership}
theory,  with  $E$ a set of possibly conditional 
equations and membership axioms,  and
$A$ a set of equational axioms such as 
associativity, commutativity, and identity, so that equational
deduction is performed \emph{modulo} the axioms $A$. 
\item $R$
is a collection of {\em labeled conditional rewrite rules\/} \( [l]:\,
t\longrightarrow t'\,\mbox{ \bf if } \mathit{cond}\), specifying  
the system's local transitions. 
\end{itemize}
A function $f$ is  declared \lstinline[mathescape]{op $f$ : $s_1\ldots s_n$ -> $s$}.
Equations and rewrite rules are introduced with, respectively,
keywords \lstinline{eq}, or  \lstinline{ceq}  for 
conditional equations, and 
\lstinline{rl} and \lstinline{crl}.
Mathematical variables 
 are declared with the keywords \lstinline{var} and \lstinline{vars}, 
 or can have the form $var\verb+:+sort$ and be introduced on the fly.

A  declaration 
\lstinline[mathescape]{class $C$ | $\mathit{att}_1$ : $\mathit{s}_1$, $\dots$ , $\mathit{att}_n$ : $\mathit{s}_n$} 
declares a \emph{class} $C$ of objects with attributes $att_1$ to $att_n$ of
sorts $s_1$ 
to $s_n$. An {\em object instance\/} of class $C$  is
represented as a term
$\texttt{<}\: O : C \mid \mathit{att}_1: \mathit{val}_1, \dots , \mathit{att}_n: \mathit{val}_n\:\texttt{>}$,
where $O$, of sort \texttt{Oid},  is the
object's
\emph{identifier}, and where $val_1$ to 
$val_n$ are the current values of the attributes $att_1$ to
$att_n$.
 A \emph{message} is a term of sort \texttt{Msg}. 
A system state
 is modeled as a term of 
the sort \texttt{Configuration}, and   has 
the structure of a  \emph{multiset} made up of objects and messages. 

The dynamic behavior of a
 system is axiomatized by specifying each of its 
transition patterns by a rewrite rule. For example, 
  the rule (with label \texttt{l})

\begin{maude}
  rl [l] :  < O : C | a1 : f(x, y), a2 : O', a3 : z >
       =>   < O : C | a1 : x + z,   a2 : O', a3 : z > .
\end{maude}

\noindent  defines a family of transitions 
in which the attribute \texttt{a1} of object \texttt{O} is updated to \texttt{x\,+\,z}. 
Attributes whose values do not change and do not affect
the next state,  
 such as \texttt{a2} and the right-hand side occurrence of \texttt{a3}, need not be mentioned. 
  %

 \paragraph{Formal Analysis in Maude.} 
Maude provides a number of  analysis methods,
including rewriting for simulation purposes, reachability analysis,
and linear temporal logic (LTL) model checking. The command
\texttt{red} $\mathit{expr}$ reduces the expression $\mathit{expr}$ to
its normal form using the equations $E$. 
The rewrite command \texttt{frew} $\mathit{init}$ simulates one
behavior from the initial state/term $\mathit{init}$ by applying
rewrite rules.  Given  a state pattern
 $\mathit{pattern}$ and an (optional) condition $\mathit{cond}$,
 Maude's  \texttt{search}
 command searches the  reachable state space from $\mathit{init}$ for
 all (or optionally a given number of) states that match
 $\mathit{pattern}$ such that 
 $\mathit{cond}$ holds:   
 
\begin{maude}
search $\mathit{init}$ =>* $\mathit{pattern}$ $[$such that $\mathit{cond}]$ .
\end{maude}


 
\paragraph{Strategies.} 
Maude provides a  language for defining strategies to control and restrict
rewriting. A strategy may not  make rewriting
deterministic, and hence multiple behaviors allowed by the strategy
must be explored. The Maude
command \lstinline[mathescape]{srew $t$ using $str$}
rewrites the term $t$ according to the strategy $str$, and   returns a
set of terms, possibly bounded by  the number of desired
solutions. \lstinline{srew} explores multiple paths 
in parallel, and ensures that
solutions will eventually be found. The
command \lstinline[mathescape]{dsrew $t$ using $str$}
 explores the behaviors allowed by $\mathit{str}$ in a depth-first
 way. 

 Basic  rewrite strategies $\mathit{str}$ include
$l$\texttt{[}$\sigma$\texttt{]} (apply  rule labeled $l$ once with the optional 
substitution $\sigma$), \code{all} (apply any of the rules, except
those marked \texttt{nonexec}, once),
\code{idle} (identity), \code{fail} (empty set), 
and \code{match $P$ s.t.\ $C$}, which  checks
whether the current term matches the pattern $P$ subject to the constraint $C$.
Compound strategies can be defined using concatenation
($\alpha\,\texttt{;}\,\beta$), disjunction ($\alpha\, \texttt{|}\, \beta$, whose result
is the union of the results of $\alpha$ and $\beta$), iteration ($\alpha \,\mathtt{*}$),
$\alpha \code{ or-else } \beta$ (execute $\alpha$, and  $\beta$
if $\alpha$ fails),  
\lstinline[mathescape]{try($\alpha$)} (applies $\alpha$ if it does not fail),
normalization  $\alpha\,\mathtt{!}$ (execute $\alpha$  until it cannot be further
applied), \texttt{matchrew} $p(x_1, \ldots, x_n)$ \texttt{s.t.}
  $\mathit{cond}$ \texttt{by}  $x_1$ \texttt{using}
    $\alpha_1$\texttt{,} \ldots\texttt{,} $x_n$
    \texttt{using} $\alpha_n$ (if the term matches the pattern
    $p(x_1, \ldots, x_n)$, then, for each match $\sigma$, rewrite each
    substitution instance 
    $x_i\sigma$ in the term according to the strategy $\alpha_i$), and
    so on~\cite[chapter 10]{maude-manual}.  

    
\paragraph{Metaprogramming.} Maude supports metaprogramming
in the sense that a Maude specification $\mathit{M}$  
can be represented as a \emph{term} $\overline{M}$ (of sort
\texttt{Module}), and a term $t$ in $M$ can be (meta-)represented as a
term $\overline{t}$ of sort \texttt{Term}.
Obtaining the meta-representations of a module $M$ and a term $t$ in
Maude is possible  
using the built-in functions \code{upModule}($M$) and
\code{upTerm}($t$) respectively. 
Maude's \texttt{META-LEVEL} module contains a number of useful
meta-level versions of key Maude functionality, including
\code{metaSrewrite} (\lstinline{srew} and \lstinline{dsrew} at the meta-level). 





\newcommand{\blaa}[1]{\textcolor{blue}{#1}}

\paragraph{Real-Time Rewrite Theories and Real-Time Maude.}
Real-time systems can be defined in rewriting logic as real-time
rewrite theories~\cite{OlvMesTCS}, which are parametric in the
(discrete or dense) time domain. In such theories, ordinary rewrite
rules model \emph{instantaneous} change, and time advance is modeled
explicitly by ``tick'' rewrite rules of the form \texttt{crl [}$\mathit{tick}$\texttt{]}
\texttt{\char123}$t$\texttt{\char125} \texttt{=>}
\texttt{\char123}$u$\texttt{\char125}  \texttt{in time} $\tau $
\texttt{if} $\mathit{cond}$, where $\tau $ is a term of sort
\texttt{Time}, $t$ and $u$ are terms of sort \texttt{System}, and
where the \emph{entire} state always has the form
\texttt{\char123}$s$\texttt{\char125} , so that \verb@{_}@ does not
occur in $s$; this ensures that time advances uniformly in the whole
system.

Real-Time Maude~\cite{rtm-journ,tacas08,wrla14} supports the modeling
and analysis of real-time rewrite theories.  Most
Real-Time Maude specifications  (not only for dense time)
have tick
rules of the form

\begin{maude}
crl [$tick$] : [$t(\overline{x})$] => {$u(\overline{x},y)$} in time $y$ if $y$ <= $f(t(\overline{x}))$ /\ $cond$ [nonexec] .
\end{maude}  

%
\noindent where $y$ is a variable 
that does not appear in $t$ and is not instantiated in
$\mathit{cond}$, making the rule non-executable 
(\texttt{nonexec}) as it stands. 

Real-Time Maude therefore offers the user the possibility of choosing between the 
following  \emph{time sampling
  strategies} for executing such \emph{time-nondeterministic} tick
rewrite rules:
\begin{itemize}
  \item \emph{deterministic time sampling}  with  a user-given
    time value $\delta > 0$;  and
    \item \emph{maximal time sampling}, with a user-given ``default'' time
      value $\delta > 0$. 
    \end{itemize}
Using \emph{deterministic} time sampling, the variable $y$ in the above tick
rule is instantiated by the selected time value $\delta$ in each
application of a time-nondeterministic tick rule; the tick rule cannot
be applied to a state \texttt{\char123}$s$\texttt{\char125} if $f(s)$
is smaller than $\delta$.

Using \emph{maximal} time sampling, the variable $y$ is 
instantiated to advance time as much as possible, namely,  
by $f(s)$ in  state
\texttt{\char123}$s$\texttt{\char125}, unless $f(s)$ is the infinity
value \texttt{INF}, in which case $y$ is instantiated with the ``default''
time value $\delta$ instead. The tick rule is not applied to state
\texttt{\char123}$s$\texttt{\char125} if the maximal time increase
$f(s)$ equals 0.

Real-Time Maude provides the following analysis methods, where the 
\emph{same}  selected time sampling 
strategy is applied in \emph{all} tick rule
applications~\cite{rtm-journ}:
\begin{itemize}
\item \emph{Rewriting} up to  time  $\Lambda$.
  \item \emph{Timed and untimed search} for  states matching a pattern
    $p(\overline{x})$, 
    such that an optional condition $\mathit{cond(\overline{x})}$
    holds,   that are reachable from the initial state
    $\mathit{init}$
    within a given  time interval $[l, u]$ (where
    $u$ could be \texttt{INF})  for time-bounded search, or in
    any time for untimed search. 
    \item \emph{Time-bounded  and  unbounded LTL model checking} check
      whether each 
      behavior from $\mathit{init}$,  up to a given time bound in the 
      time-bounded case, satisfies an (untimed) linear
      temporal logic (LTL) formula.
      \item  \emph{Timed CTL model checking}
        checks whether each behavior, possibly up to a user-given time
        bound, satisfies a given \emph{timed} CTL formula~\cite{lepri-journ}.
\item \emph{Find latest} finds the longest time it takes to reach a
  desired state.
  \item \emph{Find earliest} finds the shortest time needed to find
    the desired state.
  \end{itemize}
Since the Real-Time Maude analyses only cover the behaviors possible with
  the selected time sampling strategy, they 
   may not give the correct results~\cite{fase06}.

 In time-bounded analysis in Real-Time Maude, internally the state
 also contains the 
 ``system clock'' denoting the time it takes to reach the
 corresponding state; this adds significantly to the state space. In
 contrast, 
  in \emph{unbounded} analyses we do not carry this clock
 component with the states.

\section{The Targeted Real-Time Rewrite Theories}
 \label{sec:assumptions}

This section presents some assumptions about  the real-time rewrite
theories  we consider in the rest of this
paper.   These  assumptions 
 are needed to define  \emph{generic} timed strategies and strategy
languages. Most, if not all, large Real-Time Maude
applications naturally 
belong to this class of real-time rewrite theories, or can easily be
modified to do so (e.g., by renaming rule labels and
variables).

Section~\ref{sec:rtt} presents our running example as one such model
of a prototypical 
real-time system:  a 
simple protocol for computing the \emph{round trip times} between
pairs of senders and receivers in a network.

\subsection{Assumptions}

Since we are in the process of extending and reimplementing Real-Time
Maude, and want to use Maude features as much and directly as
possible, we specify our real-time rewrite theories directly  in Maude
by extending
the following ``timed prelude,'' which defines the sorts of our
states:

\begin{maude}
fmod TIMED-PRELUDE is   including TIME .
   sorts System GlobalSystem ClockedSystem .
   subsort GlobalSystem < ClockedSystem .
   
   op {_} : System -> GlobalSystem [ctor] .
   op _in time_ : GlobalSystem Time -> ClockedSystem [ctor] .

   var CLS : ClockedSystem .    vars T T' : Time .
   eq (CLS in time T) in time T' = CLS in time (T plus T') .
endfm
\end{maude}

We  assume a sort \texttt{Time} for the time values,
  a supersort \texttt{TimeInf} adding an infinity element
\texttt{INF} to those  values, and    assume that each tick 
rule has the form

\begin{maude}
var T : Time .  
crl [tick] : {$t$}  =>  {u} in time T  if T <= mte($t$) /\ $\mathit{cond}$ .
\end{maude}

\noindent or the form

\begin{maude}
var T : Time .  
rl [tick] : {$t$}  =>  {$u$} in time T .
\end{maude}
\normalsize

\noindent where the symbols in italics are placeholders for 
 terms and conditions.  In particular, we assume that the
 unknown
time advance is represented by the specific variable \texttt{T} (not
appearing in $t$ nor in $\mathit{cond}$), that all tick 
rules are labeled \texttt{tick}, that no non-tick rule is labeled
\texttt{tick}, and that the maximal time elapse is
given by the (user-defined) function \texttt{mte}, which  returns
a time value or \texttt{INF}.  (It is easy to define meta-level
functions renaming labels and variables to
conform to these  requirements.) 




\subsection{Running Example: Finding Round Trip Times}
\label{sec:rtt}


The following ``Real-Time Maude-style'' object-oriented Maude model
specifies 
  a very simple protocol for  computing the
\emph{round trip time} (RTT) between two nodes (a \texttt{sender} and a
\texttt{receiver})  every 5 seconds.  The delay of a message can be
any value between a lower and an upper bound. This small but
prototypical example contains many features of larger real-time distributed protocols:
clocks, timers, and messages with nondeterministic delays. 

\begin{maude}
load rtm-prelude

omod RTT is
  including TIMED-PRELUDE .
  protecting NAT-TIME-DOMAIN-WITH-INF .  

  var M : Msg .      var TI : TimeInf .      vars T T2 T3 : Time .
  vars R S : Oid .   vars C1 C2 STATE : Configuration .

  sort DlyMsg .
  subsorts Msg < DlyMsg < Configuration < System .

  op dly : Msg Time Time -> DlyMsg [ctor] .  --- upper and lower bounds

  rl [deliver] : dly(M, 0, TI) => M .    --- can deliver ripe message any time
  
  msgs rttReq_from_to_ rttResp_from_to_ : Time Oid Oid -> Msg .

  class Sender | clock : Time,  timer : Time, lowerDly : Time,  period : Time,
                 upperDly : TimeInf,  rtt : TimeInf,  receiver : Oid .
		 
  class Receiver | lowerDly : Time, upperDly : TimeInf .  

  rl [send] :
     < S : Sender | clock : T, timer : 0, period : T2,
                    lowerDly : T3, upperDly : TI, receiver : R >
    =>
     < S : Sender | timer : T2 >
     dly(rttReq T from S to R, T3, TI) .

  rl [respond] :
     (rttReq T from S to R)
     < R : Receiver | lowerDly : T3, upperDly : TI >
    =>
     < R : Receiver | >
     dly(rttResp T from R to S, T3, TI) .

  rl [recordRTT] :
     (rttResp T from R to S)
     < S : Sender | clock : T2 >
    =>
     < S : Sender | rtt : T2 monus T > .

  crl [tick] :
      {STATE}  => {timeEffect(STATE, T)} in time T
     if T <= mte(STATE) [nonexec] .

  op mte : Configuration -> TimeInf [frozen] .
  eq mte(none) = 0 .
  ceq mte(C1 C2) = min(mte(C1), mte(C2)) if C1 =/= none and C2 =/= none .

  eq mte(< S : Sender | timer : TI >) = TI .
  eq mte(< R : Receiver | >) = INF .
  eq mte(dly(M, T, TI)) = TI .
  eq mte(M) = 0 .     --- ripe message must be read immediately

  op timeEffect : Configuration Time -> Configuration .
  eq timeEffect(none, T) = none .
  ceq timeEffect(C1 C2, T) = timeEffect(C1, T) timeEffect(C2, T) 
          if C1 =/= none and C2 =/= none .
  eq timeEffect(< S : Sender | clock : T, timer : TI >, T2)
   = < S : Sender | clock : T + T2, timer : TI monus T2 > .
  eq  timeEffect(< R : Receiver | >, T) = < R : Receiver | > .
  eq timeEffect(dly(M, T, TI), T2) = dly(M, T monus T2, TI monus T2) .
  eq timeEffect(M, T) = M .

  ops snd rcv : -> Oid [ctor] .

  op init : -> ClockedSystem .
  eq init
   = {< snd : Sender | clock : 0, timer : 0, period : 5000, lowerDly : 5,
                       upperDly : 20, rtt : INF, receiver : rcv >
      < rcv : Receiver | lowerDly : 7, upperDly : 30 >} in time 0 .		       
endom
\end{maude}

Each \texttt{Sender} object has the following attributes:
\texttt{clock}    denotes its ``local clock''
(used to compute the round trip time);   \texttt{timer} denotes the
time until the next round begins; \texttt{lowerDelay} and 
\texttt{upperDelay}  give the  bounds on the delays of  messages from
the sender; \texttt{rtt} stores the last computed
round trip time value;  \texttt{period} denotes the period (e.g., five
seconds); and 
\texttt{receiver} denotes the
receiver to which it wants to know the round trip time.
A \texttt{Receiver} object only has
attributes bounding the delays of messages  \emph{from} the receiver.

A ``delayed'' message \texttt{dly(\(m\),\(\,t\),\(\,t'\))} denotes a
message $m$ whose  \emph{remaining} delay 
is in the interval $[t,t']$.  The rule  \texttt{deliver} removes the
\texttt{dly} wrapper, thereby making the message ``ripe,'' whenever
the lowest remaining delay has reached  \texttt{0}.

When the \texttt{Sender}'s \texttt{timer} expires (i.e., becomes
\texttt{0}), a new round of the RTT-finding protocol starts (rule
\texttt{send}). The
\texttt{Sender} sends an \texttt{rttReq} message with its current
clock value \texttt{T} to the \texttt{Receiver}, with appropriate
delay bounds.  The timer is also reset to expire when the \emph{next}
iteration should start.

When a \texttt{Receiver} receives such a request, it
replies with an \texttt{rttResp} message with the received
timestamp \texttt{T}, with appropriate delay bounds
(rule \texttt{respond}).

When the \texttt{Sender} receives this response, with its original
timestamp \texttt{T}, it can easily compute and store the (latest) round trip
time    (rule \texttt{recordRTT}).

The tick rule in this system, which could have many
\texttt{Sender}/\texttt{Receiver} pairs, is the usual one for
object-oriented Real-Time Maude specifications~\cite{rtm-journ}: time
can advance in a state \texttt{STATE} by any amount less than or equal
to \texttt{mte(STATE)}, and the function \texttt{timeEffect} models
how the passage of time  affects the state. \texttt{mte}
ensures that time cannot pass beyond the time when a message
\emph{must} be delivered, that time cannot pass when there is a
``ripe'' (un-delayed) message in the state, and that time cannot pass
beyond the expiration time of any \texttt{timer}. \texttt{timeEffect}
reduces the remaining  bounds of all message delays and
 timer values, and increases the \texttt{clock} values, according
to the elapsed time.

Finally, \texttt{init} defines an initial state with  one
\texttt{Sender} and one \texttt{Receiver} and the given lower and
upper bounds on the delays sent from each object, so that each
recorded RTT value should  be between 12 and 50. 

\section{Analysis Using Maude's Strategy Language Directly}
\label{sec:direct-maude-strat-lang}

This section  explains how 
 to perform ``Real-Time Maude-style''
time-sampling-strategy-based  (both time-bounded and ``clock-less''
unbounded) 
reachability analyses using Maude's strategy language,  instead of
having to use the Real-Time Maude tool.

Assuming that  the tick rule(s) are as described
above,  let $\mathit{str\myhyphen max}$ be the following strategy expression:

\small
\begin{maude}
  all | 
  (matchrew CS:ClockedSystem
     such that {STATE} in time T2 := CS:ClockedSystem /\ mte(STATE) =/= 0
       by CS:ClockedSystem
         using tick[T <- if mte(STATE) == INF then 4 else mte(STATE) fi])
\end{maude}
\normalsize

\noindent
$\mathit{str\myhyphen max}$ denotes the application of any executable
rewrite rule (\code{all}) or 
a tick rule.  If it is a \texttt{tick}
rule, then the variable \texttt{T} denoting the time increase is
instantiated to \texttt{mte(STATE)}, for the given \texttt{STATE}
obtained using \texttt{matchrew}, unless \texttt{mte(INF)} equals
\texttt{INF}, in which case \texttt{T} is set to \texttt{4}.  The tick
rule is not applied if 
\texttt{mte(STATE)} is \texttt{0}. Therefore, this expression denotes
a single application of any rule under the \emph{maximal} time
sampling strategy (with default step \texttt{4}).

\paragraph{Unbounded search} for $n$ (clocked) states matching a  \emph{state}
pattern \(\mathit{pattern}\),  with  maximal time sampling ($\mathit{str\myhyphen max}$ above), 
can be performed 
using  the  command:

\begin{maude}
    srew [$n$] $\mathit{init}$ using $\mathit{str\myhyphen max}$ *  ; (match $\mathit{pattern}$ in time T3:Time) .
\end{maude}


\begin{example}\label{ex:rtt20}
The following command checks whether it is possible to use
\emph{maximal} time sampling to find an RTT value   20:

\begin{maude}
Maude> srew [1] init using 
        $\mathit{str\myhyphen max}$ * ; (match {< snd : Sender | rtt : 20, ATTS:AttributeSet >
                            C:Configuration}  in time T3:Time) .
\end{maude}

Since the \texttt{clock} value of the sender can grow
beyond any bound, and since the desired state is not reachable with
the selected time sampling strategy, this command does not terminate.  If
we instead search for (two) reachable states where the (maximal) RTT value
50 has been recorded, we get answers:\footnote{Parts of Maude code and
  Maude output will be replaced by `\texttt{...}'.}

\begin{maude}
Maude> srew [1] init using 
        $\mathit{str\myhyphen max}$ * ; (match {< snd : Sender | rtt : 50, ATTS:AttributeSet >
                            C:Configuration}  in time T3:Time) .

Solution 1
rewrites: 589 in 0ms cpu (0ms real) (792732 rewrites/second)
result ClockedSystem:
  {< snd : Sender | clock : 50, timer : 4950, lowerDly : 5,
                    upperDly : 20, rtt : 50, receiver : rcv, period : 5000 >
   < rcv : Receiver | lowerDly : 7, upperDly : 30 >} in time 50

Solution 2
rewrites: 710 in 0ms cpu (0ms real) (734229 rewrites/second)
result ClockedSystem:
{< snd : Sender | rtt : 50, ... >  < rcv : Receiver | ... >} in time 5000
\end{maude}

\emph{Deterministic} time sampling is easier: with such time sampling
with step \texttt{1}, states with    RTT
value 20 are indeed reachable:

\begin{maude}
Maude> srew [2] init using
         (all | tick[T <- 1]) *
         ; (match {< snd : Sender | rtt : 20, ATTS:AttributeSet >
                   C:Configuration} in time T3:Time) .

Solution 1
rewrites: 18450 in 16ms cpu (16ms real) (1107310 rewrites/second)
result ClockedSystem:
  {< snd : Sender | clock : 20, timer : 4980, rtt : 20, ... >
   < rcv : Receiver | ... >}  in time 20

Solution 2
rewrites: 20409 in 18ms cpu (18ms real) (1102295 rewrites/second)
result ClockedSystem:
  {< snd : Sender | clock : 21, timer : 4979, rtt : 20, ... >
   < rcv : Receiver | ... >}  in time 21
\end{maude}          
\end{example}

\paragraph{Time-bounded Reachability Analysis.}
We can  perform \emph{time-bounded} reachability analysis to find desired
states reachable in a  time interval $[\mathit{lower},
\mathit{upper}]$ by:  (i)  applying tick rules only if the 
``system clock'' does not go beyond  $\mathit{upper}$, and (ii)  only
searching for  state patterns of clocked states whose clocks
are greater than or equal to $\mathit{lower}$.  Such time-bounded
reachability analysis should  terminate, since the tick rule is
not applied when the time bound has been reached.
%

\begin{example}
We search for states   where the desired RTT values can be found in the
time interval $[5000,10000]$.  
With maximal time sampling, we  search for two states reachable in
the desired interval where the recorded RTT value is 50:

\begin{maude}
Maude> srew [2] init using
  (all |
   (matchrew CS:ClockedSystem
      such that {STATE} in time T2 := CS:ClockedSystem /\ mte(STATE) =/= 0
          /\ T2 + (if mte(STATE) == INF then 4 else mte(STATE) fi) <= 10000
        by CS:ClockedSystem
          using tick[T <- if mte(STATE) == INF then 4 else mte(STATE) fi])) *
  ;  (match  {< snd : Sender | rtt : 50, ATTS:AttributeSet >
              C:Configuration} in time T3:Time s.t. T3:Time >= 5000) .

Solution 1
rewrites: 724 in 1ms cpu (1ms real) (498279 rewrites/second)
result ClockedSystem:
  {< snd : Sender | clock : 5000,  rtt : 50, ... >
   < rcv : Receiver | ... >}  in time 5000

Solution 2
rewrites: 749 in 1ms cpu (1ms real) (450933 rewrites/second)
result ClockedSystem:
  {< snd : Sender | rtt : 50, ... >  < rcv : Receiver | ... >
   dly(rttReq 5000 from snd to rcv, 5, 20)}  in time 5000
\end{maude}

The same   time-bounded reachability analysis terminates (with \texttt{No
  solution}) if we instead search
for a state with recorded RTT value 20.



The following time-bounded  command searches for two reachable states
with recorded RTT value 20 using deterministic time sampling (with
increment 1):

\begin{maude}
Maude> srew [2] init using
     (all | 
      (matchrew CS:ClockedSystem
         such that {STATE} in time T2 := CS:ClockedSystem
                   /\ mte(STATE) =/= 0 /\ T2 + 1 <= 10000
           by CS:ClockedSystem
             using tick[T <- 1])) *
     ;  (match {< snd : Sender | rtt : 20, ATTS:AttributeSet >
                C:Configuration} in time T3:Time s.t. T3:Time >= 5000) .

Solution 1
rewrites: 15188595 in 20685ms cpu (20835ms real) (734246 rewrites/second)
result ClockedSystem:
  {< snd : Sender | rtt : 20, ... > < rcv : Receiver | ... >}  in time 5000

Solution 2
rewrites: 15189532 in 20687ms cpu (20837ms real) (734226 rewrites/second)
result ClockedSystem: ... in time 5000
\end{maude}
\end{example}

\paragraph{Unbounded Reachability Analysis Without System Clocks.} 
\label{sec:maude-strat-for-rtm}

Our states have the form
\texttt{\char123$\mathit{state}${\char125} in time \(\mathit{clock}\)};
i.e., they include the
``system clock'' denoting how much time has passed in the system since
the execution started.  This clock is unnecessary for  \emph{unbounded}
reachability analysis, but  makes the reachable (clocked) state
space infinite, even when the (clock-less) reachable state space is
finite.

The  form of our tick rules gives us clocked states.  For
unbounded and un-clocked analysis we can either manually transform the
tick rules into the corresponding clock-less tick rules
  
\begin{maude}
crl [tick] : {$t$}  =>  {$u$} if T <= mte($t$) /\ $\mathit{cond}$ .
\end{maude}

\noindent and remove ``\texttt{in time 0}''  from the initial
states.  However,  modifying the original specification
for this single command may be undesired.

Nevertheless, the shape of our tick rules mandates that something
is done to work on clock-less states. We can extend our
modules with either the equation 

\begin{maude}
eq {STATE}  in time T = {STATE} .
\end{maude}


\noindent or with  following rule,  that removes the clock,
and always apply this rule right after applying a tick rule:

\begin{maude}
    rl [removeClock] : {STATE} in time T => {STATE} [nonexec] .
\end{maude}

We mark this rule as non-executable to allow us to control when it can be applied
(e.g., it will not be applied by the strategy \code{all}). 
\begin{example}
We can perform ``clock-less'' search for an RTT value 20 with maximal
sampling by removing the global clock after each tick:



\begin{maude}
Maude> srew [1] init using
   (all | 
    (matchrew GS:GlobalSystem
       such that {STATE} := GS:GlobalSystem /\ mte(STATE) =/= 0
         by GS:GlobalSystem
           using tick[T <- if mte(STATE) == INF then 4 else mte(STATE) fi] ; 
                 $\highlight{removeClock}$)) *
   ; (match {< snd : Sender | rtt : 20, ATTS:AttributeSet > C:Configuration}) .
\end{maude}

Unfortunately,  even the clock-less reachable
 state space is infinite in our  example, since the sender  has its own
 unbounded clock.    The same search for RTT value
50 finds the desired state.
\end{example}

\paragraph{Simulation.}  Time-bounded
\emph{simulation} of one behavior can be performed by taking one
(\texttt{one}) of the 
normal forms (\texttt{!})  when the tick rule is   
not applied if it would advance the ``system clock''  beyond the time bound:

\begin{example}
  We can simulate one behavior of our (original) RTT example up to
  time 10.000 as follows:

\begin{maude}  
Maude> srew [1] $\mathit{init}$ using
 (all |
  (matchrew CS:ClockedSystem
     such that {STATE} in time T2 := CS:ClockedSystem /\ mte(STATE) =/= 0
	 /\ T2 + (if mte(STATE) == INF then 4 else mte(STATE) fi) <= 10000
       by CS:ClockedSystem
         using tick[T <- if mte(STATE) == INF then 4 else mte(STATE) fi])) ! .

Solution 1
rewrites: 1509 in 2ms cpu (2ms real) (614163 rewrites/second)
result ClockedSystem:
  {< snd : Sender |  rtt : 50, ... > < rcv : Receiver | ... >
   dly(rttReq 10000 from snd to rcv, 5, 20)} in time 10000

\end{maude}
\end{example}




\section{A  Strategy Language for Real-Time Rewrite
  Theories}  \label{sec:core-strat}

Section~\ref{sec:direct-maude-strat-lang} shows that unbounded and
time-bounded reachability with both maximal and deterministic time
sampling can be performed using Maude's strategy
language. However, even these simple analysis methods need somewhat
hard-to-understand strategy expressions.
Furthermore, as shown in Section~\ref{sec:timed-strat}, where we
discuss strategies for real-time systems, we often need more complex
strategies. 
How can the non-Maude-expert
analyze her system with such strategies?

To address this question, in this section we 
define what we hope is 
a powerful yet intuitive timed strategy language for real-time rewrite
theories. Our focus on strategies for the non-Maude-expert is also 
motivated by 
our recent work on
providing  formal analysis and parameter synthesis for
parametric timed automata and parametric time Petri nets using Maude
with SMT solving~\cite{ftscs-journal,DBLP:conf/apn/AriasBOOPR23},
where one of our main selling
points is the ability to analyze the system with user-defined
execution strategies. However, this point becomes moot if the
timed automata/Petri net expert cannot define her strategies.

Our strategy language for real-time rewrite theories supports, e.g., 
\begin{itemize}
  \item \emph{separate} definitions of strategies for \emph{discrete
        behaviors}, including the interplay between discrete actions
      and time advance, and \emph{timed strategies}; 
  \item \emph{state-dependent time
      sampling} strategies and \emph{conditional} 
    discrete strategies; 
    \item \emph{history-dependent} strategies 
    allowing, for instance, 
    ``counting'' the number of times some states have
  been reached, manipulating such counters, and making strategies
  dependent of the values of these counters; and 
  \item intuitive syntax for ``Real-Time Maude commands''  with 
    user-defined strategies.   
  \end{itemize}

Section~\ref{sec:timed-strat} discusses important execution strategies
for both real-time systems in general and  real-time rewrite
theories, and 
  Section~\ref{sec:language} introduces our timed strategy language.
  Section~\ref{sec:semantics} 
  shows how expressions in our strategy
  language can be translated into expressions in  Maude's strategy
  language, thereby giving it a formal semantics. 
  Sections~\ref{sec:commands} and~\ref{sec:example} show how most
  Real-Time Maude analysis 
  methods can be performed using our strategy language, and
  Section~\ref{sec:benchmarking} compares the performance of our
  analysis commands with with standard Maude search on the CASH
  scheduling algorithm. 
  The executable Maude definition of our language and its semantics is
  available at~\cite{rt-strategies}.

\subsection{Strategies for Real-Time Systems}
\label{sec:timed-strat}

Interesting  execution strategies of timed systems in general include: 
\begin{enumerate}
  \item 
    \emph{Eagerness} of certain (or all)  actions:  time should not
    advance when such  actions can be taken.
  \item Advance time (or ``idle'') by $f(s)$ in all states (or
    locations) $s$
    belonging to a set of states $S$,
    advance time by $g(s')$ in all states $s'$ belonging to the set of
    states $S'$, and so on. 
\item Do not perform action $a$ more/less than $x$ times.
  \item Always execute action $a_i$ before action  $a_j$ when both are
    enabled. 
\end{enumerate}
These  examples indicate that we can consider three ``types''
of strategies: (i) strategies on the ``discrete behaviors'' (such as
items 3 and 4 above); (ii) strategies on how much to advance time
(item 2); and (iii) combining these (item 1). 
This means that the user may want to specify a
strategy restricting the discrete behaviors of a system,  as well as 
a strategy for how  to advance time.  Therefore, we
must be able
 to \emph{compose}  any \emph{discrete strategy} with  any \emph{timed
  strategy}. 

In Real-Time Maude, the selected (deterministic or maximal) time
sampling strategy  is used in all tick rule applications. However,
with maximal time sampling we may miss too many behaviors, whereas
with deterministic time sampling we may cover all possible behaviors
(for discrete time), but at the cost of ``visiting'' each time
point, even when the system is just ``idling,'' leading to \emph{very}
inefficient analyses.  In our RTT example,  the only behaviors covered
using maximal time 
sampling are those where the recorded RTT 
value is always 50. Always advancing time by 1  covers all
possible behaviors, but  requires visiting 5000 time points in
each period, even though less than 50 time points are interesting.

An efficient  time sampling strategy that covers all (interesting)
behaviors for discrete time is the following instance of item (2) above:
\begin{itemize}
\item increment time by 1 when an action \emph{could}
  happen (in the next time instant);
\item increment time maximally otherwise.
  \end{itemize}
In the RTT system, we should increment time by 1 when there is a
delayed message in the state\footnote{A further optimization would
  advance time to when the least remaining delay is 0.}, and maximally
when there is no message in the state (and the system is just 
idling until the next period begins). This suggests an 
efficient  time sampling strategy for a large class of 
distributed real-time systems~\cite{oopsla22}.

  \subsection{Our Timed Strategy Language}
  \label{sec:language}

A strategy $\langle \mu, \tau \rangle$ (of sort \code{UStrat}) in our
timed strategy language 
consists  of a user-defined 
\emph{discrete strategy} 
$\mu$  (of sort \code{UDStrat}), controlling the way instantaneous rules are
applied and their interaction with time passage, and a \emph{timed strategy} 
$\tau$ (of sort \code{UTStrat})
defining a time sampling strategy: 

\begin{maude}
sorts UStrat UTStrat UDStrat .
op <_,_> : UDStrat UTStrat -> UStrat .
\end{maude}

The discrete strategy $\mu$ controls whether some (and if
so, which) 
action/instantaneous 
rule must be applied in the current state, or whether some tick  rule must
be applied.  The timed
strategy $\tau$ defines exactly how each ``tick rule application'' (i.e., each
\texttt{delay} step) in the discrete strategy  $\mu$ is applied.

We extend the global state of the  system with a map that
 stores information  about the execution history. This
 allows us to define \emph{history-dependent strategies};   i.e., strategies
that depend on the current and the previously visited states:

\begin{maude}
sort StrState . 
pr MAP{K, V}  * (sort Entry{K, V} to Entry, sort Map{K, V} to Map) .

op _|_ : ClockedSystem Map -> StrState .
\end{maude}

\noindent The sorts \code{K} and \code{V} for the keys and their
values are user-defined.

\paragraph{Discrete Strategies.}

Discrete strategies are defined using a language whose syntax is
given as follows.  
\begin{maude}
--- Intervals 
sort Interval .
op [_,_] : Time Time -> Interval .
--- Conditions 
sort SCond . 
op matches_s.t._                : ClockedSystem Bool      -> SCond .
op matches_s.t._                : StrState Bool           -> SCond .
op matches_s.t._                : Map Bool                -> SCond .
op matches                      : ClockedSystem           -> SCond .
op in_                          : Interval                -> SCond .
ops after before after= before= : Time                    -> SCond .
ops _/\_  _\/_                  : SCond SCond             -> SCond .
op not_                         : SCond                   -> SCond .
--- User-defined strategies
op  apply_                      : Qid                     -> UDStrat .
ops  apply[_] eager[_]          : QidList                 -> UDStrat .
ops action delay eager          :                         -> UDStrat .
ops _;_  _or_  _or-else_        : UDStrat UDStrat         -> UDStrat .
op  if_then_else_               : SCond UDStrat  UDStrat  -> UDStrat .
ops stop skip                   :                         -> UDStrat .
op get_and set_                 : Map Map                 -> UDStrat .
\end{maude}

\noindent Terms of sort \code{SCond} define conditions in 
some of  the strategies. The condition
\lstinline[mathescape]{matches $P$ s.t. $C$},
where $P$ is a pattern and $C$ is an (optional)   boolean condition, 
 checks whether the current state matches $P$ so that $C$ holds in the state. 
The pattern $P$ can be a \code{ClockedSystem}, or a \code{StrState} (a clocked 
system extended with a \code{Map}). 
Other basic conditions include
checking whether the current value $t$ of the 
global clock satisfies:  $t\in [a,b]$ (\lstinline[mathescape]|in [$a$, $b$]|), 
$t > t'$ (\lstinline[mathescape]{after $t'$}),
$t \geq t'$ (\lstinline[mathescape]{after= $t'$}),
$t < t'$ (\lstinline[mathescape]{before $t'$}),  and 
$t \leq t'$ (\lstinline[mathescape]{before= $t'$}).
Larger conditions can be constructed 
using conjunction, disjunction, and negation.

User-defined  discrete strategies are: 
\lstinline[mathescape]{apply $\ell$}  applies  the
instantaneous rule with label $\ell$ \emph{once}; 
\lstinline[mathescape]{apply [$\Ell$]} applies \emph{once} the first rule
in the list of labels $\Ell$ that succeeds  in the current state
(i.e., $\Ell$ defines a \emph{priority} on the next rule to be applied);
\code{action} applies  \emph{any} instantaneous rule once; 
\code{delay} applies   a tick rule once;
\code{eager} applies the instantaneous rules as much as possible,
followed by 
\emph{one} ``delay'' when it is possible; 
\lstinline[mathescape]{eager [$\Ell$]} applies as much as possible
the rules in the list $\Ell$ followed by  one ``delay'';
 \lstinline[mathescape]{$\mu$ ; $\mu'$} is the sequential composition of
two strategies; \lstinline[mathescape]{$\mu$ or $\mu'$}
returns the union of the results obtained from the strategies $\mu$
and $\mu'$; 
\lstinline[mathescape]{$\mu$ or-else $\mu'$}  applies $\mu$, but applies
the strategy $\mu'$ 
if $\mu$ fails; 
\lstinline[mathescape]{if $\phi$ then $\mu$ else $\mu'$} is the
conditional strategy; \code{stop} is 
the strategy that always fails; 
 \code{skip} leaves the current state unchanged; and
 \lstinline[mathescape]{get $M$ and set $M'$}  uses
the pattern $M$ to retrieve  
(part of) the map storing information about the execution of the strategy 
and updates it according to $M'$.

\paragraph{Time Sampling Strategies.}
Time sampling strategies are defined as follows: 
\begin{maude}
op fixed-time_            : Time                -> UTStrat .
op max-time with default_ : Time                -> UTStrat .
sorts CTStrat LCTStrat . --- Time sampling strategies defined by cases
subsort CTStrat < LCTStrat .
op when_do_               : SCond UTStrat       -> CTStrat .
op switch_otherwise_      : LCTStrat  UTStrat   -> UTStrat .
\end{maude}

\lstinline[mathescape]{fixed-time $t$} advances the time
by time $t$
in each application of a tick rule (where advancing time by that
amount is possible). 
 \lstinline[mathescape]{max-time with default $t$}
advances time in a tick rule application by the 
maximal time $t'$ possible \emph{for that tick rule},   and advances
time by $t$ if   $t'$  is \texttt{INF}. 
%
The conditional time sampling strategy 
\lstinline[mathescape]|switch $\mathit{cases}$ otherwise $\tau$|,
where $\mathit{cases}$ is a list of choices of the form 
\lstinline[mathescape]{when $\phi_j$ do $\tau_j$},
executes the first strategy $\tau_i$ whose 
guard $\phi_i$ holds in the current state; the strategy $\tau$ is
applied if none of the guards hold.





\begin{example}\label{ex:running-example}
Consider the RTT system in Section~\ref{sec:rtt}.
    A basic  execution strategy, for any  timed strategy $\tau$, 
    applies \emph{any}  enabled
    (instantaneous or tick)  rule once: 

\begin{maude}
< delay or action , $\tau$ >
\end{maude}

It is also possible to analyze the system by assuming that instantaneous rules
have a higher priority than the tick rule: 

\begin{maude}
< eager ; $\tau$ > 
\end{maude}

In fact, we can give 
preference to the rules  \code{send} and \code{respond}, then to the
other actions, and finally 
to the tick rule: 

\begin{maude}
    < (apply ['send 'respond] or-else action or-else tick,  $\tau$ >
\end{maude}

Regarding the time sampling strategy, for any discrete strategy $\mu$, 
 the best choice for this 
system is:  if there is a \emph{delayed} message in the state,
increase 
time by 1, otherwise increase time maximally.  This
\emph{state-dependent} time sampling strategy can be defined as
follows:

\begin{maude}
< $\mu$ , switch  when matches ({CONF dly(M, T1, T2)} in time R) do fixed-time 1
      otherwise max-time with default 1 >
\end{maude}

 When the network is working well,
   we could save  bandwidth by not
  performing the RTT-finding procedure in \emph{each} period. We
  therefore add the following rewrite rule, that
   allows a \texttt{Sender} to skip a round of the protocol by
  just resetting the \texttt{timer} when it 
  expires (instead of also sending an \texttt{rttReq} message):

\begin{maude}
omod RTT-WITH-IDLING is including RTT .
 var S : Oid .    vars T T2 : Time .
  
 rl [skipRound] :
   < S : Sender | timer : 0, period : T2 > => < S : Sender | timer : T2 > .
endom
\end{maude}

\noindent When its timer expires, a sender  therefore
  nondeterministically chooses between executing a round of the
  protocol (rule \texttt{send}) or skipping one round (rule
  \texttt{skipRound}).

A sensible strategy  is to skip some rounds but never 
 skip more than two rounds in a
row. 
To define this \emph{state- and history-dependent} strategy, we use a counter
labeled with \lstinline|'C| 
to avoid skipping ``more than two rounds'':  

\begin{maude}
< delay or 
  if matches {< S : Sender | timer : 0, ATTS >} in time T  --- State dep. 
  then if (matches ('C |-> N) s.t. N <= 1) --- History dependent 
       then apply 'skipRound ; 
            (get ('C |-> N) and set ('C |-> N + 1))  --- Skip and increment 
       else apply 'send ; 
            (get ('C |-> N) and set ('C |-> 0)) --- Send and reset
  else action ) , $\tau$ >
\end{maude}
\end{example}

\subsection{Semantics}\label{sec:semantics}
This section  shows how expressions in our timed strategy language
 can be translated into expressions in  Maude's
strategy language. Hence, the denotational and  operational semantics
of the latter \cite{maude-strategy-language} formally describes the
execution of real-time rewrite theories controlled by  a timed
strategy $\langle \mu, \tau \rangle$. 

We define a map $\enc{-}$ from
terms of sort \code{UStrat} to terms of sort \code{Strategy}, the sort in Maude's
prelude used to meta-represent strategies. 

\begin{figure}
\begin{subfigure}{\textwidth}
        \small
        \[
        \begin{array}{l}
            \enccond{\code{matches SS  s.t.  B}}  =  \enccond{\code{matches } \overline{\code{SS}} \code{ s.t. } \overline{\code{B}} }\\
            \enccond{\code{matches } \code{CS} \code{ s.t. B} }  =  \enccond{\code{matches  CS | M s.t. } \overline{\code{B}} }\\
            \enccond{\code{matches } \code{M} \code{ s.t.  B} }  =  
            \enccond{\code{matches } \code{ $\overline{\code{CS | (M , M')}}$ s.t. } \overline{\code{B}} }\\
            \enccond{\code{matches  Te s.t. Te'}}  =  \code{match Te s.t. Te' }\\
            \enccond{\code{after(T)}} = \enccond{\code{match  \{ CS \}  in time  T'  s.t.  T > T'}}\\
        \enccond{\phi_1 \wedge \phi_2 } = \enccond{\phi_1} \code{ ; } \enccond{\phi_2}\qquad
        \enccond{\phi_1 \vee \phi_2 } = \enccond{\phi_1} \code{ or-else } \enccond{\phi_2}\\
        \enccond{\code{not } \phi} = \code{not } \enccond{\phi} 
    \end{array}
\]
\caption{Conditions. Definitions for \code{before}, \code{in}, etc are similar and omitted.  \label{fig:bool}}
\end{subfigure}
\begin{subfigure}{\textwidth}
        \small
        \[
        \begin{array}{lll}
            \enctime{\code{fixed-time  T1}}  =  \code{'tick [ $\overline{\code{T}}~\leftarrow ~\overline{\code{T1}}$  ] \{ empty \}}\\
            \enctime{\code{max-time with default T1}}  =  
            \code{matchrew } \overline{\code{SS}} \code{ s.t. } \overline{\code{ \{ S \}  in time  T2 | M}} \code{ := } \overline{\code{SS}} 
            \\
            \multicolumn{3}{l}{
            ~~~\code{by } \overline{\code{SS}} \code{ using } \code{'tick [} \overline{\code{T}} \leftarrow \overline{\code{if INF == mte(S)} \code{ then }  \code{T1} \code{ else } \code{mte(S)} \code{ fi} }  \code{ ] \{ empty \}}} \\
            \enctime{\code{switch } (\code{when C  do } \tau)~\code{LC}  \code{ otherwise } \tau' }  =  \enccond{\code{C}}~\code{ ? } \enctime{\tau} \code{ : } \enctime{\code{switch } \code{LC} \code{ otherwise } \tau' }\\
            \enctime{\code{switch } (\code{when C  do } \tau)  \code{ otherwise } \tau' }  =  \enccond{C}~\code{ ? } \enctime{\tau} \code{ : } \enccond{\tau'}
    \end{array}
\]
 \caption{Timed strategies. } \label{fig:tstrat}
\end{subfigure}

\begin{subfigure}{\textwidth}
        \small
        \[
        \begin{array}{lll}
            \multicolumn{3}{l}{
                \enc{\la \code{stop } , \tau\ra}  =  \code{fail}  \qquad
                \enc{\la \code{skip } , \tau\ra} =  \code{idle} \qquad 
            \enc{\la \code{apply Q} , \tau\ra}  =  \code{Q [none] \{empty\}}}\\
            \multicolumn{3}{l}{
    \enc{\la \code{action} , \tau\ra}  =  \code{all} \qquad
            \enc{\la \code{delay} , \tau\ra}  =  \enctime{\tau} \qquad
    \enc{\la \code{eager} , \tau\ra}  =  \code{ all ! ; try(} \enctime{\tau} \code{) } } \\
        \enc{\la \code{apply [nil]} , \tau\ra}  =  \code{fail} \quad
        \enc{\la \code{apply [Q LQ]} , \tau\ra}  =  \code{apply Q or-else } \enc{\code{apply [LQ]}} \\
        \enc{\la \code{eager [L]} , \tau\ra}  =   \enc{\la \code{apply [}L\code{]}  ,\tau \ra} \code{ ! : try(} \enctime{\tau} \code{)}\\
            \enc{\la \mu \code{ ; } \mu', \tau\ra}  =  \enc{\la \mu, \tau \ra} \code{ ; } \enc{\la \mu', \tau \ra} \qquad 
            \enc{\la \mu \code{ or } \mu', \tau\ra}  =  \enc{\la \mu, \tau \ra} \code{ | } \enc{\la \mu', \tau \ra}\\
            \enc{\la \mu \code{ or-else } \mu', \tau\ra}  =  \enc{\la \mu, \tau \ra} \code{ or-else } \enc{\la \mu', \tau \ra}\\
            \enc{\la \code{if  C  then } \mu \code{ else } \mu', \tau\ra}  =  \enccond{\code{C}} \code{ ? } \enc{\la \mu,\tau \ra} \code{ : } \enc{\la \mu',\tau \ra} \\
            \enc{\la \code{get M' and set  M''}, \tau  \ra }\quad = 
            {\code{matchrew } \overline{\code{SS}} \code{ s.t. } \overline{\code{\{ S \} in time  T1 |  (M, M')}} \code{ := } \overline{\code{SS}} } \\
            \multicolumn{3}{l}{~~~\code{by } \overline{\code{SS}} 
                \code{ using 'updateMap [ } \overline{\code{M}} \leftarrow \overline{\code{M}} \code{ ; }
                \overline{\code{M'}} \leftarrow \overline{\code{M'}} \code{ ; }
                \overline{\code{M''}} \leftarrow \overline{\code{M''}} 
                \code{ ] \{ empty \}}
                }
    \end{array}
\]
\caption{Discrete and real-time strategies.} \label{fig:tstart}
\end{subfigure}
\begin{subfigure}{\textwidth}
        \small
        \[
        \begin{array}{l}
            \enc{\la \code{check } \phi, \tau \ra }  =  \enccond{\phi} \qquad \qquad\qquad
            \enc{\la \code{until } \phi \code{ do } \mu, \tau \ra }  = (\enccond{\phi} \code{ ? fail : } \enc{\la\mu,\tau\ra}) \code{! } \\
            \enc{\la \code{repeat }  \mu, \tau \ra }  = \enc{\la\mu,\tau\ra}) \code{ * } \qquad\ 
            \enc{\la \code{0} \code{ steps with  }  \mu, \tau \ra }  = \code{idle} \\
            \enc{\la \code{s(N)} \code{ steps with  }  \mu, \tau \ra }  = \enc{\la\mu,\tau\ra}) \code{ ; } {\la \code{N  steps with  }  \mu, \tau \ra }\\
            \enctime{\code{untime } \tau} = \enctime{\tau} \code{ ; 'removeClock [ none ] \{ empty \}}
    \end{array}
\]
\caption{General timed strategies.  \label{fig:enc-ext}}
\end{subfigure}

\caption{Interpretation of real-time strategies as Maude's strategies.\label{fig:interpretation}}
\end{figure}


\begin{definition}[Semantics]
The interpretation of conditions ($\enccond{-}$), time sampling strategies 
($\enctime{-}$), and real-time strategies ($\enc{-}$), as terms of sort
\code{Strategy} is given in Fig. \ref{fig:interpretation}. 
These definitions use the following variables, and require the new operator and rule below: 
\begin{maude}
vars M M' M'' : Map . var CS : ClockedSystem . var SS : StrState . 
var B : Bool .  vars Te Te' : Term . vars  T T' T1 T2 : Time . 
var C : SCond . var LC : LCTStrat . var S : System . 

op matching_s.t._ : Term Term -> SCond .
rl [updateMap] : CS | (M, M') => CS | (M, M'') [nonexec] .
\end{maude}
\end{definition}

In Figure \ref{fig:interpretation}, 
we use $\overline{t}$ to denote the meta-representation
of a term $t$ (in Maude, \code{upTerm($t$)}). 
For instance, the second case in Figure \ref{fig:bool}
    must be read as 
   \[
       \enccond{\code{matching CS s.t.  B} }  =  \enccond{\code{matching  `_|_[}\overline{\code{CS}}, \code{`M:Map ] s.t. } \overline{\code{B}} }
   \]
   and specified in Maude as 
   \begin{maude}
eq enc(matching CS s.t. B) = 
   enc(matching '_|_[upTerm(CS), 'M:Map] s.t. upTerm(B))
   \end{maude}

The Maude strategy $\enccond{\phi}$ fails when condition $\phi$ does not hold, 
and succeeds (without modifying the current state) otherwise. 
\code{matches} expressions are reduced until their
parameters  are of sort \code{Term}. Then, 
Maude's strategy \code{match} is used to check whether the current
state matches the pattern 
and satisfies the given condition (otherwise, \code{match} fails). 

The Maude strategy $\enctime{\tau}$ applies the tick rule by instantiating 
the variable \code{T} with the needed substitution according to $\tau$. 
In \code{max-time}, Maude's strategy \code{matchrew}
is used to do pattern matching and bind the variable \code{S}
with the current configuration. Hence, the call 
\code{mte(S)} determines the next tick value. 
The  definition of  \code{switch} uses the conditional
Maude strategy \lstinline[mathescape]{$\alpha$ ? $\beta$ : $\gamma$}
to choose the right time sampling strategy $\tau_i$. 

The Maude strategy $\enc{\langle \mu,\tau\rangle}$ fails when $\mu=$
\code{stop} and does nothing if $\mu=$\code{skip}. If $\mu=$\code{ apply Q},
the rule with label \code{Q} is applied, without any substitution (\code{[none]})
and with the \code{\{empty\}} list of strategies (since no particular strategy
is used to solve rewrite expressions in conditional rules). Maude's strategy \code{all} 
non-deterministically chooses, and applies
once, any of the \emph{executable} rewrite rules.
Therefore,  when $\mu=$ \code{action},
only executable instantaneous (and no tick) rules are applied. The strategy
$\enc{\code{apply [}\Ell \code{]}}$ tries, in order, the instantaneous rules in the
list $\Ell$. In the case $\mu=$\code{delay}, the strategy
$\enctime{\tau}$ is executed. The normalization operator \code{all !} applies
\code{all} until it cannot be further applied. Hence, when $\mu=$ \code{eager},
all the instantaneous transitions are (non-deterministically) 
taken and then, if it is possible, a tick is performed. 
The
interpretation of the strategies \code{_;_}, \code{_or_}, \code{_or-else_} and
\code{if_then_else} uses the corresponding constructors in Maude's strategy
language. In the case $\mu=${\code{get M' and set M''},
  Maude's \code{matchrew} 
is used to bind $M'$ with the needed entries in the map storing information 
about the execution of the strategy. Then, 
the execution of the rule 
\code{updateMap} replaces the values in  $M'$ with the corresponding ones in $M''$. 

The ``extended'' discrete and timed strategies implementing
the different analyses proposed in Section \ref{sec:commands} 
are defined as Maude's strategies as shown in Figure \ref{fig:enc-ext}.
The Maude's strategy $\enc{\la \code{check } \phi~,~\tau\ra}$ 
executes $\enccond{\phi}$ (thus failing when $\phi$ does not hold).
The Maude's strategy 
$\enc{\la \code{until } \phi \code{ do } \mu~,~\tau \ra}$ uses the normalization operator to repeatedly 
execute $\enc{\la \mu,\tau\ra}$ until the point where  $\enccond{\phi}$ succeeds.
The iteration operator \code{*} is used to give meaning to the 
strategy 
\lstinline[mathescape]{repeat $\mu$}, that iteratively executes  $\mu$ until it
fails. The Maude's  strategy $\enc{\la n \code{ steps with } \mu~,~\tau\ra}$ 
stops executing $\mu$ when $n=0$. Otherwise, it is recursively
called after  applying $\enc{\la \mu, \tau\ra }$. 
Finally, the strategy $\enctime{\code{untime } \tau}$ executes $\enctime{\tau}$
and then applies the rule \code{removeClock}. 

\subsection{User-Friendly Analysis Commands}
\label{sec:commands}

A user-defined strategy $\langle \mu, \tau \rangle$ controls ``one round'' of
the execution of the system.   In this section we provide convenient
``Real-Time Maude-like'' syntax for most simulation,
reachability and other formal analysis methods provided by Real-Time
Maude, albeit executed with user-defined strategies.

We define new discrete and time sampling
strategies (sorts \code{DStrat}, \code{TStrat} and \code{Strat}) that control
how user's strategies are applied. Building on these new strategies, 
different analysis for real-time rewrite theories can be neatly defined. 

Discrete strategies (of sort \code{DStrat}), besides the basic
user-defined discrete strategies, include:  
the strategy \lstinline[mathescape]{check $\phi$} that fails if $\phi$
does not hold in the current state; the conditional repetition of a given
strategy \lstinline[mathescape]{until $\phi$ do $\mu$}; the strategy
\lstinline[mathescape]{repeat $\mu$} that iteratively executes  $\mu$ until it
fails; and the  strategy \lstinline[mathescape]{$n$ steps with $\mu$} that
repeats $n$ times $\mu$. Overloaded operators for the sort \code{DStrat} (e.g.,
\lstinline{ op _;_ ... [ditto]}) are also defined and omitted here. 

\begin{maude}
sort DStrat .   subsort UDStrat < DStrat .
op check_                : SCond                  -> DStrat .
op until_do_             : SCond DStrat           -> DStrat .
op repeat_               : DStrat                 -> DStrat .
op _steps with_          : Nat DStrat             -> DStrat .
\end{maude}

General timed strategies (of sort \code{TStrat}) extend user-defined
time sampling strategies with a new case, used later to define untimed reachability
analysis: 
\begin{maude}
sort TStrat .    subsort UTStrat < TStrat . 
op untime  : TStrat -> TStrat .
\end{maude}

\noindent $\code{untime } \tau$ applies 
$\tau$ and then 
the rule \code{removeClock}, thus
removing the global clock from the current state. 
These new strategy constructors are defined as Maude strategies
as shown in Figure 
\ref{fig:enc-ext}.


\paragraph{Commands.}
We  now define a convenient syntax for most Real-Time Maude-like analysis
commands using strategies. 
Given a user-defined strategy $\la \mu,\tau \ra$, we define an extended
strategy $\la \mu',\tau' \ra$ that implements such an analysis command  by
rewriting (using Maude's \code{metaSrewrite}) an initial state
$\mathit{init}$   and returning  
a list of \code{ClockedSystem}s (the solutions). 

Time-bounded simulation is implemented with the following command:
\begin{maude}
op tsim [_]  in_:_using_with sampling_until_ :  
   Nat Qid StrState  DStrat TStrat Time -> LClockedSystem .
\end{maude}


The command \lstinline[mathescape]|tsim [$n$] in $R$ : $\mathit{init}$ using $\mu$ with sampling $\tau$ until $r$|
returns  the first $n$ states that result when  
rewriting $\mathit{init}$ 
in theory $R$ when following the strategy 
$\la (\code{until after=(}r\code{) do } \mu)  , \tau\ra$.
In words, this strategy returns the first reachable states, in each branch of the 
search tree, whose 
global clock $t$ 
satisfies $t\geq r$. The parameter \code{[n]} is optional. 

It is also  possible to observe the behavior of the system up to  a given number $d$ of rewriting steps: 
\begin{maude}
op trew [_,_]  in_:_using_with sampling_ :  
  Nat Nat Qid StrState  DStrat TStrat -> LClockedSystem .
\end{maude}

The command \lstinline[mathescape]|trew [$d$, $n$] in $R$ : $\mathit{init}$ using $\mu$ with sampling $\tau$|
returns the first $n$ states that can be reached after $d$ 
rewriting steps. For that, the initial state is rewritten with the 
strategy 
$\la d \code{ steps with } \mu)  , \tau\ra$.

Unbounded and time-bounded reachability analyses are implemented by the
following commands:  
\begin{maude}
op tsearch [_] in_:_=>_using_with sampling_ :  
   Nat Qid StrState SCond DStrat TStrat -> LClockedSystem .
op tsearch [_] in_:_=>_using_with sampling_in time_ :  ... -> ... .
\end{maude}

\noindent \lstinline[mathescape]|tsearch [$n$] in $R$ : $\mathit{init}$ => $\phi$ using $\mu$ with sampling $\tau$|
returns the first $n$ states that result from $\mathit{init}$ by rewriting with 
the  strategy 
$\la \code{repeat } \mu \code{ ; check } \phi, \tau\ra$, i.e., 
repeat $\mu$  zero or more times and, on the resulting term,
check $\phi$. 
Time-bounded reachability analysis
 \lstinline[mathescape]|tsearch [$n$] in $R$ : $\mathit{init}$ => $\phi$ using $\mu$ with|\newline
\lstinline[mathescape]|sampling $\tau$ in time [$a$,$b$]|
 is implemented as the  extended strategy 

\noindent$\la \code{repeat (} \code{if after}(b) \code{ then stop else } \mu \code{) ; check } (\phi \wedge \code{in [} a,b\code{]}), \tau\ra$.

``Depth-bounded'' versions of the form 
\lstinline[mathescape]|tsearch [$n$,$d$] in $R$...|
of the above commands
are available; they 
check whether a $\phi$-state can be reached by 
applying at most $d$ times the strategy $\mu$ 
(\lstinline[mathescape]{$d$ steps using $\mu$} \code{|}
\lstinline[mathescape]{$d-1$ steps using $\mu\cdots$)}.
Furthermore, similar commands \code{dsearch} are defined where 
\code{metaSrewrite} is invoked  with the flag \code{depthFirst}, thus 
exploring the rewriting graph in depth.

%

Untimed reachability analysis is possible with the command 
\begin{maude}
op usearch [_] in_:_=>_using_with sampling_ :  
   Nat Qid StrState SCond DStrat TStrat -> LClockedSystem .
\end{maude}

The implementation of this command is similar to the one for \code{tsearch} but
the sampling strategy used is \code{untime(}$\tau$\code{)}: 
after each tick, the global clock is removed from the state. A
depth-first version \code{dusearch}  is also available.

Finding the longest and the shortest time it takes to reach a desired
state is supported by    
the following commands:

\begin{maude}
op find latest in_:_=>_using_with sampling_ :  
    Qid StrState SCond DStrat TStrat -> LClockedSystem .
op find earliest in_:_=>_using_with sampling_ :  ... -> LClockedSystem .

\end{maude}

 \code{find latest} uses \code{metaSrewrite} to
find \emph{all} the solutions when rewriting the initial state   with the
 strategy $\la \code{until } \phi \code{ do } \mu \code{ ; check } \phi
~,~\tau\ra$. 
This  finds the \emph{first} state
in \emph{all} the branches of the search tree that satisfies $\phi$. 
We then post-process the returned list to find the state  with the greatest 
global clock value. This command may not terminate if there is a
branch where $\phi$ never holds. 

The command \code{find earliest} cannot be implemented with a procedure similar
to the one for \code{find latest}. The reason is that \code{find earliest} must
finish if there is a reachable state satisfying  $\phi$, even if there are
branches  not leading to $\phi$-states.  We therefore 
compute the first solution  when applying the strategy $\la \code{until } \phi \code{
do } \mu \code{ ; check } \phi ~,~\tau\ra$. Let $t$ be  the global
clock value in
this state. 
The strategy $\la
\code{until } \psi \code{ do } (\code{if after}(t) \code{ then stop else } \mu)
\code{ ; check } \psi ~,~\tau\ra$,
where $\psi$ is the condition $\phi \wedge
\code{before}(t)$, is then used to find a new solution (if any)
whose global clock is strictly smaller than $t$. This
branch-and-bound 
procedure is repeated until no further solutions are found.

 \subsection{Example: Analyzing the Round Trip Time Protocol}
\label{sec:example}


 We illustrate the use of our timed strategy language on the RTT
 example.

 We  check whether an RTT value 20 can be recorded when  incrementing 
 time by 1 in each tick rule application:\footnote{Symbols in all
   capitals are variables, whose 
   declarations we often omit.}
\begin{maude}
Maude > red tsearch [2] in 'RTT : init => 
        matches ({CONF < S : Sender | rtt : 20, ATTS >} in time R:Time) 
        using delay or action with sampling fixed-time 1 .

result NeList{ClockedSystem}: 
  ({< snd : Sender | rtt : 20, ... > < rcv : Receiver | ... >}  in time 20) 
  ({< snd : Sender | rtt : 20, ... > < rcv : Receiver | ... >}  in time $\highlight{21}$)
\end{maude}

It is possible to reach two states with  RTT 50 using maximal time
sampling:
\begin{maude}
Maude > red tsearch [2] in 'RTT : init => 
        matches ({CONF < S : Sender | rtt : 50, ATTS > } in time R:Time ) 
        using delay or action with sampling max-time with default 4 .

result NeList{ClockedSystem}: 
 ({< snd : Sender | clock : 50, timer : 4950, rtt : 50, ... > 
   < rcv : Receiver | lowerDly : 7, upperDly : 30 >}  in time 50) 
 ({< snd : Sender | clock : 5000, rtt : 50, ... > 
   < rcv : Receiver | lowerDly : 7, upperDly : 30 >}  in time $\highlight{5000}$)
\end{maude}

As already shown in Example \ref{ex:rtt20}, it is not possible to 
find an RTT value 20 when the maximal time-sampling strategy is used. 
Consider the 
\emph{state-based} sampling strategy in Example \ref{ex:running-example},
where time advances by 1 when there are delayed messages, and maximally otherwise:
\begin{maude}
< action or delay , 
  switch  when matches ({CONF dly(M, T1, T2)} in time R) do fixed-time 1
          otherwise max-time with default 1 >
\end{maude}

This strategy allows us to find states with RTT value 20 while visiting
less states when compared to the \code{fixed-time} strategy. 
Note, for instance, the global clock of the second solution found by
the command below, 
and compare it with the outputs in the two commands above: 

\begin{maude}

Maude> tsearch [2]  in 'RTT : init => matches ...
       using delay or action with sampling  $\tau$ .

result NeList{ClockedSystem}: 
  ({< snd : Sender | rtt : 20, ... >  ... } in time 20) 
  ({< snd : Sender | rtt : 20, ... >  ... } in time $\highlight{5000}$)
\end{maude}

We can also search for a state with RTT value 50 reachable in the time
interval $[5000,10000]$ with maximal time sampling:
\begin{maude}
Maude> red tsearch [2] in 'RTT : init => matches ... using delay or action 
       with sampling max-time with default 4 in time [5000, 10000] .

result NeList{ClockedSystem}: 
  ({< snd : Sender | rtt : 50, ... > }  in time 5000) 
  ({< snd : Sender | rtt : 50, ... > dly(...)}  in time 5000)
\end{maude}
The same search for an RTT value 20 yields no solution:
\begin{maude}
Maude> red tsearch [2] in 'RTT : init => matches ... using delay or action 
       with sampling max-time with default 4 in time [5000, 10000] .

result LClockedSystem: (nil).LClockedSystem  --- (No solution)
\end{maude}
Untimed and ``clock-less'' search for RTT value 50:
\begin{maude}
Maude> red usearch [2] in 'RTT : init => matches ...
       using delay or action with sampling max-time with default 4 .

result NeList{ClockedSystem}: 
  {< snd : Sender | rtt : 50, ... > < rcv : Receiver | ... >}  
  {< snd : Sender | rtt : 50, ... > < rcv : Receiver | ... >}
\end{maude}
Time-bounded simulation  simulates one system behavior  up to time 10000:
\begin{maude}
Maude> red tsim [1] in 'RTT : init
       using delay or action with sampling max-time with default 4 until 10000 .

result ClockedSystem: {< snd : Sender | rtt : 50, ... > 
                       < rcv : Receiver | lowerDly : 7, upperDly : 30 >}  
                       in time 10000
\end{maude}
We then check the longest and shortest time needed to record an RTT
value different from \texttt{0} and \texttt{INF}
(\texttt{rtt?(STATE)}) for the first time in
each behavior:
\begin{maude}
Maude> red find earliest  in 'RTT : init => matches ({ STATE} in time T2) 
       s.t. (rtt?(STATE)) using action or delay with sampling fixed-time 1 .

result ClockedSystem: {< snd : Sender | rtt : 12, ... > ... }  in time 12

Maude> red find latest  in 'RTT : ... with sampling fixed-time 1 .

result ClockedSystem: {< snd : Sender | rtt : 50, > ... }  in time 50

Maude> red find latest  ... using $\highlight{eager}$ with sampling fixed-time 1 .

result ClockedSystem: {< snd : Sender | rtt : 12, ... > ... }  in time 12
\end{maude}


Finally, let $\mu$ be the history-depend strategy in Example
\ref{ex:running-example}. The execution of the following commands
 show that some states
are not visited when this is strategy is followed, resulting in a smaller state space.  
(\code{size(L)} returns the size of the list \code{L}).

\begin{maude}
Maude> red size(tsearch  in 'RTT : init => matches ({STATE} in time T2) 
       using $\mu$ with sampling max-time ...  in time [0, 100000]) .

result NzNat: 126

Maude> red size(tsearch  in 'RTT : init => matches ({STATE} in time T2) 
       using action or delay with sampling max-time ... in time [0, 100000]) .

result NzNat: 162
\end{maude}

\subsection{Benchmarking}
\label{sec:examples}  \label{sec:benchmarking}

We have proposed and implemented a strategy language and shown that
all (except for temporal logic model checking at the moment) Real-Time
Maude analysis can be performed using strategies, with much more
flexibility in defining both discrete strategies and time sampling
strategies.   The question is
whether such Maude-strategy-based implementation of reachability
analysis has competitive performance, compared to standard Maude
reachability analysis.

We therefore compare the performance of our implementation with
standard Maude search on a Maude model of a  Real-Time Maude
benchmark systems, a variation of the CASH scheduling algorithm
developed by Marco Caccamo at UIUC~\cite{original-cash}.
The idea of the CASH is that some jobs may not
need all the execution time allocated to it.   These unused clock
cycles are put in a \emph{queue}, so that other jobs could use  them
to improve system performance.   CASH is a sophisticated protocol,
with sporadic tasks 
(i.e., a job could arrive at any time), unknown length of each job,
and a 
queue of unused execution times. 

Real-Time Maude analysis in~\cite{fase06} showed, somewhat
surprisingly, that the length of the queue of unused execution times could
grow beyond any bound. This  means that most formal real-time tools cannot
analyze this version of  CASH.  Real-Time Maude
analysis  discovered that a hard deadline could be
missed~\cite{fase06}. 
This was a subtle (and previously unknown) flaw: the deadline miss was
not found by extensive 
randomized simulation. 

    We transformed the Real-Time Maude  specification of the CASH
    protocol into a ``standard''  Maude model by   incrementing
     time by one unit in the tick rules (see~\cite{rt-strategies}).
    We can therefore use Maude's  \lstinline{search} command
    to find whether it is possible to 
    reach a state where a deadline is missed within time
    12:\footnote{Here \texttt{init} denotes an initial state from
      which a missed deadline should not be reachable if the optimized
      version of CASH were correct.}
    \begin{maude}
Maude> search [1] init =>* {DEADLINE-MISS CONF} in time T s.t. T <= 12 .

Solution 1 (state 599272)
rewrites: 34093729 in 14910ms cpu (14937ms real) ...
... 
\end{maude}

The   \code{tsearch} and \code{dtsearch} (depth-first
search) commands in our language that correspond  
to this time-bounded reachability query
are executed as follows:   

\begin{maude}
Maude> red tsearch [1] in 'CASH : init => 
           matches ({DEADLINE-MISS CONF} in time T ))
           using delay or action with sampling fixed-time 1 in time [0, 12] .

rewrites: 44 in 19517ms cpu (19538ms real) ...

Maude> red dtsearch [1] ... in time [0, 12] .

rewrites: 44 in 2079ms cpu (2083ms real) ...
\end{maude}

For a possibly fairer comparison, we  also perform \emph{unbounded}
reachability analysis with 
Maude \code{search},  \code{tsearch}, and \code{usearch};
\code{dtsearch} failed to terminate in this case.

We first apply Maude's \code{search} command, without constraints on the
system clock:

\begin{maude}
Maude> search [1] init =>* {DEADLINE-MISS C:Configuration} in time T:Time .

Solution 1 (state 599272)
states: 599273  rewrites: 34093728 in 14988ms cpu (15010ms real) ...
\end{maude}  

And then compare it to our own unbounded \code{tsearch} command and
our \code{usearch} command, which always applies a rule which removes
the system clock after each tick step: 

\begin{maude}
Maude> red tsearch [1] ... .
rewrites: 36 in 21906ms cpu (21951ms real) ...

Maude> red usearch [1] ... .
rewrites: 39 in 9778ms cpu (9790ms real) ...
\end{maude}  

For an even more optimized Maude \code{search}, we can 
modify our Maude specification by \emph{manually} removing the
``\texttt{in time ...}'' part 
of each each tick rule, so that the state does not carry the system
clock:

\begin{maude}
search [1] init =>* {DEADLINE-MISS CONF} . 

Solution 1 (state 151069)
states: 151070  rewrites: 8601350 in 4498ms cpu (4503ms real) ...
\end{maude}
%
%
%





 All  experiments were run on a Dell XPS 13 laptop (with an
 Intel i7 processor @ 1.30GHz and 16GB of RAM). 
 For time-bounded reachability analysis,   our \code{tsearch} command
 (20 seconds runtime) 
is not much slower than Maude's \code{search} command 
(15 seconds) on a Maude
model where the deterministic time sampling strategy with increment 1  is
hard-coded in the tick rule.   Furthermore, our ``depth-first'' search
command \code{dtsearch}  significantly outperforms Maude's
search command on this application  (2 seconds).

In the unbounded case, it is fair to compare \code{tsearch} (22
seconds), which carries the system clock, to the Maude \code{search}
which took 15 seconds, and \code{usearch} (9.8 seconds), which
explicitly removes the clock after each tick,  to Maude \code{search} of
the manually modified model which never introduces the system clock
(4.5 seconds).

To summarize, our strategy-implemented commands are not much 
more than twice as slow as Maude \code{search}, and in one case even
significantly faster.


Because of the high degree of time-nondeterminism in the model (jobs
can arrive at any time and may execute for an arbitrary amount of
time), the only time sampling strategy that makes sense for CASH is
deterministic with increment 1.  The performance gain should be much larger on
systems such as RTT where we should use mixed time sampling to
``pass'' idling states where not much can happen.  Although these
preliminary results are quite promising, we should perform more
thorough benchmarking in future work.

 \section{Related Work}
\label{sec:related}

\paragraph{Strategies for real-time systems.}
{\sc Uppaal Stratego}~\cite{uppaal-stratego} extends the timed
automaton tool {\sc Uppaal}~\cite{uppaal} with strategies and model
checking under such strategies, where startegies are {\sc Uppaal}
queries. {\sc Uppaal Stratego} seems mainly to be used in connection
with machine learning-based synthesis of controller strategies.   We
target a more expressive model, provide a language for specifying
actual strategies instead of queries, and must also provide time
sampling strategies. We also perform model checking w.r.t.\ a
strategy, but do not support synthesizing  strategies. 

\paragraph{Rewrite strategies.} Different strategy languages have been proposed
to cope with the inherent non-determinism in rewriting systems. 
Examples of such languages
include ELAN \cite{DBLP:journals/ijfcs/BorovanskyKKR01}, Stratego
\cite{DBLP:journals/scp/BravenboerKVV08}, and $\rho$Log
\cite{DBLP:journals/jancl/MarinK06}. 
Applications of Maude's strategy
language~\cite{maude-strategy-language} include the analysis of
different systems and models such as neural 
network \cite{DBLP:conf/dcai/Santos-GarciaPV08} and membrane systems
\cite{DBLP:journals/jlap/RubioMPV22}, as well as  the specification of
semantics of programming languages
\cite{DBLP:journals/entcs/Hidalgo-HerreroVO07} and process calculi
\cite{DBLP:journals/entcs/VelardoSV06}.

Our previous work
\cite{ftscs22,DBLP:conf/apn/AriasBOOPR23,ftscs-journal} showed, for the first time,  how
Maude strategies
can be used on simple real-time systems. That work motivated
the development of an easy-to-use timed strategy language, which
resulted in this paper. 

Recently, Rubio et al.
\cite{DBLP:journals/jlap/RubioMPV21,DBLP:journals/ase/RubioMPV22} have shown
how to model check strategy-aware rewriting logic specification. These
techniques have been implemented in the \code{umaudemc} tool, that allows for
model checking LTL and CTL formulas, as well as to perform probabilistic and
statistical model-checking on systems controlled by strategies. This will open
the possibility of model checking real-time rewrite theories following the
real-time strategies proposed here.


The paper~\cite{beffara2009verification} uses rewrite rules and
``strategies'' to analyze timed automata reachability using the
rewriting framework
ELAN~\cite{DBLP:journals/ijfcs/BorovanskyKKR01}. In this setting, the
authors 
define data types and rewrite rules for manipulating ``zones'' of the
timed automaton, and then define rewrite strategies for various approaches to
analyze these symbolic state spaces of the automaton.  In other words:
they use rewrite strategies not to restrict the possible behaviors of
the timed automaton, but to define various analysis methods on the
graph of zones of all behaviors of the automaton.  In contrast, we
define strategies to define and explore certain subsets of behaviors
of timed systems.

\section{Concluding Remarks}  \label{sec:concl}
In this paper we propose what we hope is a  useful yet reasonably
intuitive language for defining execution strategies for real-time
systems in Maude, allowing to perform most of the analysis methods
supported by Real-Time Maude,  with user-defined strategies. Our
language allows the user to define her discrete strategies and her
time sampling strategies separately.   We identify a number of
interesting execution strategies for real-time systems, including a
``mixed'' time sampling strategy (not supported by Real-Time Maude)
that should be ideal for explicit-state analysis of a large class of
distributed real-time systems, such as our round trip time
protocol. 

Our strategies are given a semantics in Maude and are therefore
implemented in Maude. A preliminary performance comparison between
standard Maude 
search and our strategy-implemented reachability analyses  on
the CASH scheduling algorithm benchmark indicates that the latter are
fairly competitive.

The benefits of this work are: (i) allowing the user to quickly and
easily analyze her real-time system  under a wide range of
different scenarios without having to perform the error-prone task of
modifying her model; (ii) providing much ``better'' (and flexible)
time sampling strategies for time-sampling-based explicit-state
analysis than provided by Real-Time Maude; (iii) providing a
convenient framework in which we can quickly experiment with different
strategies and analyses, before optimizing and hard-coding the most
promising into the Real-Time Maude tool; (iv) allowing us to analyze
real-time rewrite theories directly in Maude, instead of in Real-Time
Maude, which is currently being redeveloped; and (v) providing a
backend for analyzing with user-defined strategies  also 
time automata and time Petri nets, as well as for other modeling
languages and formalisms for which Real-Time Maude provides a formal
analysis backend.

Much work remains. For example, we should also provide untimed and timed temporal
logic model checking combined with real-time strategies.
For that, the \code{umaudemc} tool \cite{DBLP:journals/jlap/RubioMPV21}, and its support for 
model checking LTL and CTL formulas on strategy-controlled rewrite theories, 
will be fundamental. 
%
We also plan to  combine
\emph{symbolic} analysis of real-time rewrite theories with
user-defined strategies, as preliminary we have done in
\cite{ftscs22} and \cite{DBLP:conf/apn/AriasBOOPR23}.

\paragraph{Acknowledgments.} The authors acknowledge support from 
the NATO Science for Peace and Security Programme through grant number
G6133 
(project SymSafe) and the 
PHC project Aurora AESIR.



\end{document}